\documentclass[prb,
superscriptaddress,showpacs,amsmath,amssymb]{revtex4}

\begin{document}

\title{Extended Tsallis-Cirto entropy for black and white holes}

\author{G.E.~Volovik}
\affiliation{Landau Institute for Theoretical Physics, acad. Semyonov av., 1a, 142432,
Chernogolovka, Russia}

\date{\today}

\begin{abstract}
In reference [3] we considered the black hole thermodynamics with the non-extensive entropy. This entropy obeys the composition rule which coincides with the composition rule in the non-extensive Tsallis-Cirto $\delta=2$ statistics. Here we extend this approach to the thermodynamics of white holes. The entropy of the white hole is negative as follows from the rate of macroscopic quantum tunneling from black hole to white hole. The white hole entropy is with the minus sign the entropy of the black hole with the same mass, 
$S_{\rm WH}(M)=-S_{\rm BH}(M)$. This reflects the anti-symmetry with respect to time reversal, at which the shift vector in the Arnowitt-Deser-Misner formalism changes sign. This symmetry allows one to extend the Tsallis-Cirto entropy by adding a minus sign to the Tsallis-Cirto formula applied to white hole. As a result, the composition rule remains the same, with the only difference being that instead of entropy it contains the entropy modulus. The same non-extensive composition rule is obtained for the entropy of the Reissner-Nordstr\"om black hole. This entropy is formed by the positive entropy of the outer horizon and the negative entropy of the inner horizon. The model of the black hole formed by "black hole atoms" with Planck-scale mass is also extended to include the negative entropy of white holes.
\end{abstract}
\pacs{
}

\maketitle

\tableofcontents

\section{Introduction. Composition rule for black hole entropy}

The thermodynamics of black holes can be obtained by studying quantum tunneling processes. Quantum tunneling of particles from a black hole determines the Hawking temperature.\cite{Wilczek2000} On the other hand, macroscopic quantum tunneling, which controls the processes of dividing a black hole into smaller parts, determines the black hole entropy.
\cite{Volovik2022}  

The black hole entropy $S_{\rm BH}(M)=4\pi GM^2$ is non-extensive with the special type of composition.  In the process of the splitting of the black hole with mass $M$ into two smaller black holes with masses $M_1$ and $M_2$, with $M_1+M_2=M$, the entropy obeys the following rule:
\begin{equation}
\sqrt{S_{\rm BH}(M=M_1 +M_2)}= \sqrt{S_{\rm BH}(M_1)} +\sqrt{S_{\rm BH}(M_2)}\,.
\label{BlackHoles}
\end{equation}
The entropy decreases after splitting, $S_{\rm BH}(M_1) + S_{\rm BH}(M_2)<S_{\rm BH}(M_1 +M_2)$. The process of splitting followed by decrease of entropy occurs in the process of macroscopic quantum tunneling with probability:\cite{Volovik2022}  
\begin{eqnarray}
w \propto \exp(-\Delta S) =\exp\left[-S_{\text{BH}}(M)+S_{\text{BH}}(M_1)+S_{\text{BH}}(M_2)\right] \,.
\label{tunnelingM1M2}
\end{eqnarray}
Equation (\ref{tunnelingM1M2}) demonstrates that the quantum process of splitting is equivalent to a thermodynamic fluctuation whose probability is expressed through the difference in entropy before and after the splitting. The inverse process of merging of two black holes occurs with increase of the entropy and does not require thermodynamic or quantum fluctuations.

As discussed in Ref. \cite{Volovik2025a}, the composition rule in Eq.(\ref{BlackHoles}) is consistent with the non-extensive Tsallis-Cirto $\delta=2$ entropy:\cite{TsallisCirto2013,Tsallis2020}
\begin{equation}
S_{\delta =2}=\sum_i p_i \left(\ln\frac{1}{p_i} \right)^2\,,
\label{TCentropy}
\end{equation}
which gives
\begin{equation}
\sqrt{S_{\delta =2}(A+B)}=\sqrt{S_{\delta =2}(A)} + \sqrt{S_{\delta =2}(B)}\,.
\label{TCentropy2}
\end{equation}
Application of the Tsallis-Cirto $\delta=2$ statistics to black hole was also considered in Ref.\cite{Manoharan2025}. 

The goal of this paper is to modify the Tsallis-Cirto $\delta=2$ statistics in order to incorporate the white holes and also the black holes with two horizons. In Section \ref{WHSec} the modified Tsallis-Cirto $\delta=2$ entropy appropriate for white hole is introduced. The general composition rule for processes where both black and white holes participate is in Section \ref{CombinedSec}. Application to the Reissner-Nordstr\"om black hole, in which the inner horizon has negative entropy, is in Section \ref{RNsection}.

Section \ref{AtomsSec} is devoted to a toy model which reflects the modified Tsallis-Cirto $\delta=2$  thermodynamics of black and white holes. In this model, black and white holes are formed from elementary building blocks — elementary holes with Planck-scale mass. 
Section \ref{NakedSec} discusses the discontinuous transition of the Reissner-Nordstr\"om black hole to extremality as topological Lifshitz transition.

\section{Tsallis-Cirto entropy for white hole}
\label{WHSec}

The white holes are the configurations with the opposite shift vector in Arnowitt-Deser-Misner formalism.\cite{ADM2008}  In the Painleve-Gullstrand coordinates \cite{Painleve,Gullstrand}  the metric has the following form:
\begin{equation}
ds^2= - dt^2(1-{\bf v}^2) - 2dt\, d{\bf r}\cdot {\bf v} + d{\bf r}^2 \,,
\label{PGmetric}
\end{equation}
where for the Schwarzschild black and white holes the shift function is:
\begin{equation}
{\bf v} ({\bf r})=\mp \hat{\bf r} \sqrt{\frac{2MG}{r}}\,.
\label{velocity}
\end{equation}
In condensed matter analogs, this vector corresponds to the velocity of the "quantum vacuum" or, in Ellis's drainhole analogy \cite{Ellis1973}, the velocity of the "ether". The minus sign in the equation (\ref{velocity}) corresponds to the black hole configuration (the velocity ${\bf v} $ is directed toward the horizon), and the plus sign describes a white hole (the "ether" flows away from the horizon).

The entropy of the white hole can be found from the calculated rate of the macroscopic quantum tunneling from the black hole with mass $M$ to the white hole with the same mass.\cite{Volovik2022} 
The obtained tunneling exponent is:
\begin{equation}
w\propto  \exp{\left(-2S_\text{BH}(M)\right)}
\,.
\label{tunneling}
\end{equation}
As in Eq.(\ref{tunnelingM1M2}), we can consider the process of quantum tunneling as thermodynamic fluctuation, $w\propto  \exp(-\Delta S)= \exp(S_\text{WH}-S_\text{BH})$. Then one obtains that entropy of the white hole with mass $M$ is with minus sign the entropy of the black hole with the same mass:
\begin{equation}
 S_{\rm WH}(M)=-S_{\rm BH}(M) =-4\pi GM^2
\,.
\label{WHBH}
\end{equation}
This equation reflects the antisymmetry between black and white holes with respect to time reversal.
 Negative entropy of a white hole confirms the statement, that white hole is the extremely fine-tuned state.\cite{Wittem2025,Eardley1974} 

Interestingly, in the condensed matter simulations the situation can be exactly the opposite.
The white-hole horizon for the “relativistic” ripplons at the surface of the shallow water is easily simulated using the kitchen-bath hydraulic jump, see Fig. 4 in \cite{Volovik2006}. Here the role of the shift function ${\bf v}$ is played by the fluid flow. This white hole does not decay because the flow of water through the horizon is compensated by the influx from the central singularity formed by the vertical jet of water. On the other hand, the reverse process requires extremely fine-tuning, and the corresponding black hole is highly unstable.

Since the black and white hole entropies in Eq.(\ref{WHBH}) differ only by sign, this antisymmetry suggests the antisymmetric extension of the Tsallis-Cirto $\delta=2$ entropy for the white hole, i.e. it  should be the Equation (\ref{TCentropy}) with minus sign:
\begin{equation}
S^-_{\delta =2}=-\sum_i p_i \left(\ln\frac{1}{p_i} \right)^2\,.
\label{TCentropyMinus}
\end{equation}
In the next section we show that such an extension yields composition rules that properly include white holes in the processes of fission and merging.

\section{Combined composition rule}
\label{CombinedSec}

With Eq.(\ref{TCentropyMinus}), the composition rule in Eq.(\ref{BlackHoles}) can be easily extended to describe processes that also involve white holes with their negative entropy:
\begin{equation}
\sqrt{|S(M_1 +M_2)|}= \sqrt{|S(M_1)|} +\sqrt{|S(M_2)|}\,.
\label{BWHoles}
\end{equation}
This includes in particular the composition rule, which describes the  process of quantum tunneling from the black hole to white hole:
\begin{equation}
\sqrt{|S_{\rm WH}(M)|}= \sqrt{|S_{\rm BH}(M)|}\,.
\label{EqualHoles}
\end{equation}
This equation is consistent with the anti-symmetry between black and white holes in Eq.(\ref{WHBH}).

The macroscopic Schwarzschild hole cannot be partially black and partially white. It is either black or white. But in the processes of splitting, the black or white hole can split into the holes of both types.
The black hole can split into two smaller black holes:
\begin{equation}
S_{\rm BH}(M=M_1 +M_2) \rightarrow  S_{\rm BH}(M_1) +S_{\rm BH}(M_2) =4\pi G(M_1^2+M_2^2) <4\pi GM^2,
\label{Splitting1}
\end{equation}
or into the black-white pair:
\begin{eqnarray}
S_{\rm BH}(M=M_1 +M_2) \rightarrow  S_{\rm BH}(M_1) +S_{\rm WH}(M_2)=4\pi G(M_1^2-M_2^2)<4\pi GM^2,
\label{Splitting2}
\end{eqnarray}
or into two white holes:
\begin{eqnarray}
S_{\rm BH}(M=M_1 +M_2) \rightarrow  S_{\rm WH}(M_1) +S_{\rm WH}(M_2)=-4\pi G(M_1^2+M_2^2) <4\pi GM^2\,.
\label{Splitting3}
\end{eqnarray}
All the above processes of splitting of the black hole are described by the composition rule in Eq.(\ref{BWHoles}). This composition rule shows that in all of these processes, the entropy decreases, as it does in rare thermodynamic fluctuations.\cite{Landau_Lifshitz} 

The process of macroscopic quantum tunneling, in which a black hole transforms into a white hole, is a thermodynamic fluctuation in which the entropy decreases to its most negative value at a fixed mass $M$. The further processes of the splitting of the white hole are always accompanied by increase of entropy from its most negative value. The increase of entropy is the natural process in thermodynamics, which obeys the second law of thermodynamics. This shows that it is the negative entropy which is responsible for the second law of thermodynamics for the white hole. Together with antisymmetry, the second law of thermodynamics of a white hole justifies the emergence of negative entropy.

\section{Composition rule for Reissner-Nordstr\"om black hole}
\label{RNsection}

The modified composition rule in Eq. (\ref{BWHoles}) can be applied for the calculation of the entropy of Reissner-Nordstr\"om (RN) black hole.  The entropy of the RN black hole can be considered as the composition of the entropies of the outer and inner horizons.
Entropy of the outer horizon is
\begin{equation}
S_{\rm RN}(r_+)= \pi r_+^2/G=\pi G \left(M+ \sqrt{M^2 -\alpha Q^2/G} \right)^2\,,
\label{outerRN}
\end{equation}
where $\alpha$ is the fine structure constant and $Q$ is the integer valued electric charge with $Q=-1$ for electron. 
The entropy of the inner horizon is negative:
 \begin{equation}
S_{\rm RN}(r_-)= -\pi r_-^2/G=-\pi G \left(M- \sqrt{M^2 -\alpha Q^2/G} \right)^2\,.
\label{innerRN}
\end{equation}
The minus sign for the entropy of the inner horizon in Eq. (\ref{innerRN}) is due to the fact that the inner horizon is equivalent to the white hole horizon and therefore has negative entropy. The negative entropy of the inner horizon agrees with the negative temperature of the inner horizon, $T_-= - (r_+ -r_-)/4\pi r_-^2$, see Eq.(53) in Ref. \cite{Volovik2022}.
 
Application of the composition rule in Eq. (\ref{BWHoles}) suggests that the total entropy of the RN black hole has the following value:
 \begin{equation}
S_{\rm RN}(M)= \left( \sqrt{|S_{\rm RN}(r_+)|} + \sqrt{|S_{\rm RN}(r_-)|}\right)^2= \pi(r_+ +r_-)^2/G =4\pi GM^2\,.
\label{entropyRN}
\end{equation}
According to Eq.(\ref{entropyRN}) the total entropy of the RN black hole does not depend on the conserved electric charge $Q$ and is completely determined by the mass of the black hole. Thus the RN black hole has the same entropy as the Schwarzschild black hole with the same mass $M$. This is not surprizing, since the RN black hole can be adiabatically converted into the Schwarzschild black hole by continuously decreasing the fine structure constant $\alpha$ to zero at fixed $M$ and $Q$.\cite{Volovik2022} Such adiabatic transformation does not change the entropy of the black hole.

Note that in Ref. \cite{Volovik2022} we considered the entropy of the inner horizon with the sign +.  Although this is not correct, fortunately it does not affect the equation (\ref{entropyRN}), since the latter does not depend on the sign of the entropy of the inner horizon.

The composition rule (\ref{entropyRN}) can be generalized to the arbitrary number of the horizons. 
Let us consider the multi-horizon metric similar to that in Ref.\cite{Odintsov2017}. For the black hole with $n$ horizons in Minkowski background we first choose the simplest metric with
 \begin{equation}
-g_{00}(r)=1-{\bf v}^2(r)= \frac{1}{r^{n}}\prod_{a=1}^{n} (r-r_a)\,\,,\,\, \sum_{a=1}^n r_a= 2MG\,.
\label{metricmulti}
\end{equation} 
The extension of the composition rule (\ref{entropyRN}) giives the following total entropy for this black hole with its alternating black and white horizons with correspondingly positive and negative entropies: 
\begin{equation}
S= \left( \sum_{a=1}^{n} \sqrt{|S(r_a)|}  \right)^2=\frac{\pi}{G}\left(\sum_{a=1}^{n}r_a\right)^2 =4\pi GM^2\,.
\label{entropymulti}
\end{equation}
Entropy again depends only on the mass of the black hole.
Since this result is consistent with the adiabatic transformation of a multi-horizon black hole into a Schwarzschild black hole, it is valid for the general multi-horizon metric in Minkowski background.

\section{Atoms of black and white holes}
\label{AtomsSec}

The Tsallis-Cirto $\delta=2$ statistics inspires the toy model, in which the mass spectrum of the black hole is equidistant, $M=Nm_{\rm P}$, where $N$ is an integer number and $m_{\rm P}=1/\sqrt{8\pi G}$ is the reduce Planck mass. In this model, the black hole can be considered as ensemble of $N$  black hole "atoms" with Planck mass scale.\cite{Volovik2025P}
The modified Tsallis-Cirto entropy allows us to extend the consideration to the white hole. 
In the toy model, the black hole "atom" has zero entropy. Thus according to the anti-symmetry in Eq.(\ref{WHBH}), the white hole "atom" also must have zero entropy. This is why we can simply identify a white hole "atom" with a black hole "atom", assuming that these quanta are indistinguishable. Then according to Ref.\cite{Volovik2025P}
we have
\begin{equation}
S_{\rm BH}(N) =\frac{N(N-1)}{2}= -S_{\rm WH}(N)\,.
\label{BWHolesN2}
\end{equation}
This gives zero entropy for a single "atom", $S(N=1)=0$.
On the other hand in the thermodynamics limit, i.e. when the number of "atoms" $N\gg 1$, the splitting of black hole with $N$ "atoms" to the smaller black holes and the splitting of the white hole to the smaller white holes obey the following composition rule:
\begin{equation}
\sqrt{|S(N=N_1 +N_2)|}= \sqrt{|S(N_1)|} +\sqrt{|S(N_2)|}\,.
\label{BWHolesN}
\end{equation}

The negative entropy of a white hole reflects the effort expended in preparing such a state. The rate of splitting of the black hole with into $N$ "atoms" is $w\propto \exp(-N(N-1)/2)$. In this process the entropy decreases from $S_{\rm BH}(N)=N(N-1)/2$ to $S=0$. The formation of the white hole from $N$ "atoms" is also extremely unlikely, and can be achieved by quantum fluctuation with the rate $w\propto \exp(-N(N-1)/2)$. That is why, in the process of constructing a white hole from “atoms”, the entropy decreases even more, now from $S=0$ to the negative value $S_{\rm WH}(N)=-N(N-1)/2$. 

As we already discussed in Sec. \ref{WHSec}, the equation $S_{\rm BH}(N) = -S_{\rm WH}(N)$ reflects the anti-symmetry with respect to time reversal, at which the shift vector ${\bf v}$ of PG coordinates changes sign, $T{\bf v}=-{\bf v}$, and the black hole transforms to the white hole. The time reversal of the process of formation of a black hole is equivalent to the process of relaxation of a white hole. The decrease in entropy of black hole in this inverse process corresponds to the increase in entropy during the relaxation of a white hole. As a white hole relaxes, its entropy increases, starting from the most negative value, in agreement with the second law of thermodynamics.

 In the same manner the time reversal transforms the expanding de Sitter with the positive Hubble parameter, $H>0$, to contracting de Sitter with $H<0$. Correspondingly, due to anti-symmetry of the Hubble parameter, $TH=-H$, the entropy of the volume inside the cosmological horizon is negative in the contracting Universe, $S(H<0) =-S(H>0)=-A/4G$, where $A$ is the area of the cosmological horizon.\cite{Volovik2025b} 
 
 \section{Extremality and naked singularity}
\label{NakedSec}

In all the cases considered in previous sections, negative entropy occurs in the presence of a horizon: this is the horizon of a white hole, the horizon in a contracting de Sitter, and the inner horizon of a black hole (see also Refs. \cite{Odintsov2002,Odintsov2017}). A natural question arises: can negative entropy take place in thermodynamic systems without horizons? 

When described in terms of an ensemble of Planck-scale quanta ("planckons"\cite{Aharonov1987,YenChinOng2025}), the entropy of black and white holes depends only on the number $N$ of quanta. Therefore, one might expect that the equation (\ref{BWHolesN2}) for the entropy holds for the RN black hole also in super-critical regime, where $\alpha >G M^2/Q^2$. In this regime, there are no horizons, but the entropy can be concentrated in a naked singularity. This is consistent with the claim that all the entropy of an ordinary Schwarzschild black hole is concentrated in the black hole singularity.\cite{Volovik2025k}
The naked singularity has been discussed in various contexts, e.g., particle production in a naked singularity was considered in \cite{Henheik2025}; primordial naked singularities were discussed in \cite{Joshi2025,Joshi2024}; whether naked singularities can be distinguished from black holes by observations was discussed in \cite{Virbhadra2022}; etc.
Thus, it is not excluded that the positive and negative entropies in equation (\ref{BWHolesN2}) can be obtained in the supercritical regime by an ensemble of "planckons" concentrated in a naked singularity. 
Let us consider the possible scenarios.

The transition to a super-critical black hole by continuously changing the fine structure constant $\alpha$ is non-adiabatic, as can be seen from the entropy $S_+$ and $S_-$ of outer and inner horizons:
\begin{equation}
S_\pm  =\pm  \frac{N(N-1)}{8}\left(1 \pm \sqrt{1-8\pi\alpha q^2}  \right)^2 \,\,, \,\, q=\frac{Q}{N} \,.
\label{RNentropues}
\end{equation}
Although equation (\ref{BWHolesN2}) can be obtained by continuously (adiabatically) varying $\alpha$ to zero, one cannot be sure that equation (\ref{BWHolesN2}) will be valid in the supercritical regime, $\alpha >1/8\pi q^2$. The approach to extremality is not continuous,\cite{Carroll2009} and the entropy may jump from $N(N-1)/2$ to zero when two horizons merge and annihilate. In the same manner the entropy of the RN white hole will jump from its negative value $-N(N-1)/2$ to zero when horizons annihilate. This brings us back to the situation where the composition rule and negative entropy only hold in the presence of horizons.

If the entropy of a black hole is concentrated at the singularity, then it would be difficult to justify its jump to zero when the two horizons merge. Instead, one can assume that equation (\ref{RNentropues}) can be extended to the supercritical regime, where the horizons are in the complex plane. In this case, the composition rule in Eq.(\ref{entropyRN}) may suggest that the entropy of the horizonless black and white holes, which is concentrated in the naked singularity, is:
\begin{equation}
S_{\rm BH} = -S_{\rm WH}=\left( \sqrt{|S_{\rm RN}(r_+)|} + \sqrt{|S_{\rm RN}(r_-)|}\right)^2= 8\pi \alpha Q^2\,\,, \,\, \alpha>\frac{N^2}{8\pi Q^2}\,.
\label{SuperBWHoles}
\end{equation}
In this case, the super-extreme black hole is unstable with respect to splitting into a pair: sub-extreme + super-extreme black holes. Another problem with this scenario is that in the supercritical regime the entropy depends on $\alpha$, and thus the adiabatic principle does not hold. 

So we return to the scenario with zero entropy of a supercritical black hole. The zero entropy can be justified in the following way. In the absence of a horizon, the thermodynamic temperature can be determined only by the Planck scale singularity, and thus the temperature may have the Planck energy scale, $T\sim m_{\rm P}$.
Then the entropy of the singularity is $S\sim M/T \sim M/m_{\rm P}=N$, which is much smaller than the entropy $N(N-1)/2$ of a subcritical black hole with horizons. In the thermodynamic limit, $N\rightarrow \infty$,
this corresponds to a zero entropy, $S=0$. Moreover, in this scenario the adiabatic principle also becomes valid: in the supercritical regime the entropy $S=0$ does not depend on the parameter $\alpha$. The adiabatic principle is violated only at the transition point between the two regimes at $\alpha=1/8\pi q^2$, when the horizons are formed. The event horizon is the source of irreversibility, which provides positive entropy for a black hole and negative entropy for a white hole.

 The discontinuous transition can be considered as a type of topological quantum phase transitions, in which topological invariants change discontinuously.\cite{Volovik2007} In our case, horizons are formed, destroyed, annihilated, or change their topology.\cite{Jacobson1995,Smolin2003} 
The corresponding topological invariant can be provided by Euler characteristic.\cite{Liberati1997} The horizon can be considered as the space-time analogue of the Fermi surface, which undergoes Lifshitz transitions,\cite{Lifshitz1960} while singularity is the analogue of the Weyl point in the spectrum of fermionic quasiparticles. The Weyl point can be isolated, corresponding to a naked singularity, or enclosed by the Fermi surface.\cite{VolovikZhang2017,Volovik2018}

Because of antisymmetry, a supercritical white hole should also have zero entropy. 
 In the  supercritical regimes the black hole and its antisymmetric clone are the same physical objects. This is in contrast to the subcritical regime, where the coordinate transformation between the black and white hole PG metrics in the equation (\ref{PGmetric}) has singularities at the horizons. That is why the general covariance does not apply, and the subcritical black and white holes are different physical objects with opposite entropies. The singular coordinate transformation between these objects provides another way to calculate the tunneling rate from a black hole to a white hole.\cite{Volovik2022}

Let us mention some papers from the recent conference "Black Holes Inside and Out 2024", which have connection to this topic. These are: the paper on possibility of the entropy concentrated in the black hole singularity; \cite{Wald2024} the paper on possible instability of the extremal black hole and on the extremality as a precursor to naked singularities;\cite{Dafermos2024} and the paper by Unruh, which concludes that the quantum mechanics of black holes is still a far from finished topic.\cite{Unruh2024}

\section{Conclusion}
\label{ConclusionSec}

We modified the non-extensive Tsallis-Cirto $\delta=2$ statistics to include into consideration the white holes, which have negative entropy. We used the anti-symmetry between the black and white hole, due to which the entropy of white hole is with minus sign the entropy of the black hole with the same mass, $S_{\rm WH}(M)=-S_{\rm BH}(M)$. This anti-symmetry takes place because under the time reversal the shift vector ${\bf v}$ in the Arnowitt-Deser-Misner formalism changes sign, $T{\bf v}=-{\bf v}$,  thus transforming the black hole to white hole. This anti-symmetry allows one to extend the Tsallis-Cirto entropy by adding a minus sign to the Tsallis-Cirto equation, when it is applied to white hole. As a result,  the modified composition rule contains the entropy modulus instead of entropy. 

The anti-symmetry  provides the negative entropy to all the white-type horizons, including the cosmological horizon of the contracting de Sitter state and the inner horizon of the Reissner-Nordstr\"om black hole. For the RN black hole, the modified composition rule combines the positive entropy of the outer horizon and the negative entropy of the inner horizon. This rule shows that the total entropy of an RN black hole is completely determined by its mass $M$ and does not depend on the conserved (integer) charge $Q$. This is consistent with the adiabatic transformation of the RN black hole into a Schwarzschild black hole, where the parameter $\alpha$ (the fine structure constant) adiabatically transforms to zero for fixed $M$ and $Q$. During this adiabatic transformation, the entropy does not change. This concerns also the multi-horizon black holes with alternating black and white horizons in Minkowski background.

The toy model, in which the black hole entropy is formed by black hole "atoms" has been also extended to include white holes. Modified Tsallis-Cirto $\delta=2$ statistics showed that the corresponding "atom"  with Planck-scale mass has zero entropy. These neutral elementary quanta serve as the building blocks of both black holes with positive entropy and white holes with negative entropy.

\end{document}